# Enlightening the atomistic mechanisms driving self-diffusion of amorphous Si during annealing

Iván Santos*, Luis A. Marqués, Lourdes Pelaz
*Departamento de Electricidad y Electrónica, Universidad de Valladolid,E.T.S.I. de Telecomunicación, Paseo Belén 15, 47011 Valladolid, Spain*

Luciano Colombo
Dipartimento di Fisica, Università degli Studi di Cagliari, Cittadella  Universitaria, I-09042 Monserrato (CA), Italy

ABSTRACT

We have analyzed the atomic rearrangements underlying self-diffusion in amorphous Si during annealing using tight-binding molecular dynamics simulations. Two types of amorphous samples with different structural features were used to analyze the influence of coordination defects. We have identified several types of atomic rearrangement mechanisms, and we have obtained an effective migration energy of around 1 eV. We found similar migration energies for both types of samples, but higher diffusivities in the one with a higher initial percentage of coordination defects.

Despite the renewed interest in amorphous Si (*a*-Si) for optoelectronic and photovoltaic applications [1], the atomic-scale understanding of its microstructural evolution during processing (an issue of large technological impact) is still a matter of debate. In particular, it has been found that impurity diffusion is greatly affected by modifications in the *a*-Si structure [2, 3]. These modifications mainly occur during annealing, and they are directly related to microscopic changes of the amorphous matrix [4]. Uncontrolled diffusion during the fabrication process results in the degradation of the final performance of devices. A better understanding of diffusion at atomic level will help to overcome these negative effects.

In this brief report we have analyzed the atomic rearrangements underlying self-diffusion in *a*-Si during annealing. We have used tight binding molecular dynamics (TBMD) simulations within an *sp$^3$* orthogonal tight-binding approach, and the hamiltonian representation proposed by Kwon *et al.* [5]. This representation has been successfully used for studying diffusion phenomena in crystalline Si [6-8] (*c*-Si) and the properties of *a*-Si [9, 10]. We have considered two models of *a*-Si, obtained by rather different computer simulations. One the one hand, *a*-Si was obtained by melting *c*-Si, and quenching the resulting liquid at a rate of $10^{15}$ K/s, which lyes within the typical cooling rates used in molecular dynamics simulations [11]. On the other hand, *a*-Si was obtained using the algorithm developed by Wooten, Winer and Weaire (WWW) [12]. We considered three quenched and two WWW cubic cells with 216 atoms. The structural differences between these two types of *a*-Si simulation cells are well known: while the WWW algorithm generates a perfectly four-fold coordinated *a*-Si [12], quenched samples have an appreciable occurrence of non four-fold coordinated atoms that depends on the quench rate [11]. In our case we had ~ 80% and ~ 20% of four- and five-fold coordinated atoms, respectively. The aim of considering two different models of *a*-Si is to investigate the possible role of coordination defects on the microstructure evolution during annealing. Generated samples were rapidly heated at the desired temperature and stabilized during 5 ps before analyzing the dynamics of the system. Then, we performed constant temperature and volume simulations for total times ranging from 1.5 to 4 ns. The temperature range was chosen between 900 and 1300 K, since for higher temperatures *a*-Si transforms into liquid [9]. We performed velocity rescalings every $10^3$ time steps to keep the temperature of the cell constant. Equations of motion were integrated using the velocity-Verlet algorithm [13] with a time step of 1.5 fs. Periodic boundary conditions were applied in all directions.

To analyze self-diffusion, we evaluated the atomic mean-squared displacement (MSD) as

$$MSD(t) = \sum_{i=1}^{N} \frac{|\vec{r_i}(t) - \vec{r_i}(0)|^2}{N}$$

**Equation 1**

being $N$ the number of atoms in the simulation cell, and $\vec{r_i}(t)$ the position of atom $i$ at time $t$. A representative plot of the time evolution of the MSD in the case of quenched $a$-Si at several temperatures is shown in Figure 1. At temperatures lower than 1000 K, the atomic MSD slowly increases with time by means of small steps separated by long plateaus. As temperature increases, the steps are higher and plateaus become shorter. Bearing in mind that abrupt changes in the MSD have been correlated to transitions among different configurations of the Si self-interstitial in $c$-Si [14], we paid attention to atomic trajectories when abrupt changes in the MSD occurred in some of the simulations performed. We used the Fast Fourier Transformation (FFT) to filter out thermal vibrations for identifying relevant atomic movements in our simulations. The FFT is applied to atomic trajectories to obtain the frequency spectrum. Since frequencies associated to lattice vibrations are much higher than those associated to atomic rearrangements, an inverse FFT in which high frequencies are neglected is performed to obtain filtered atomic trajectories and keep only those events that really induce microstructural changes. Using this procedure we could identify different types of atomic rearrangements, namely: *bond break* (BB), *bond switch* (BS), *frustrated bond switch* (FBS), *kick-out* (KO), and the *vacancy like* (VL), which are represented in Figure 2. These rearrangements should not be considered as unique and well-defined mechanisms but, rather, as `prototypical mechanisms' since the underlying amorphous network makes the number of different atomic configurations virtually infinite. Contrary to point defect diffusion in $c$-Si, once a rearrangement occurs in $a$-Si a new atomic environment not equivalent to the initial configuration is obtained. Analogous atomic rearrangements may happen following a similar path (but not exactly equal) to those that occurred previously. This results in a distribution of energy barriers for each type of mechanism rather than a unique value, as it was also observed in theoretical studies of the configurational energy landscape in $a$-Si [15].

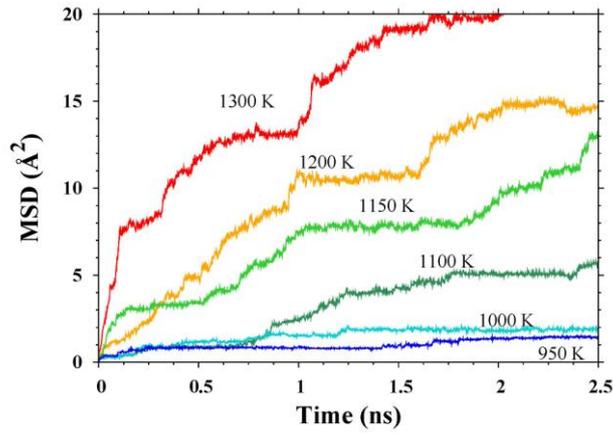

**Figure 1** (Color online) Representative plot for the MSD in the case of quenched *a*-Si as a function of time for different temperatures.

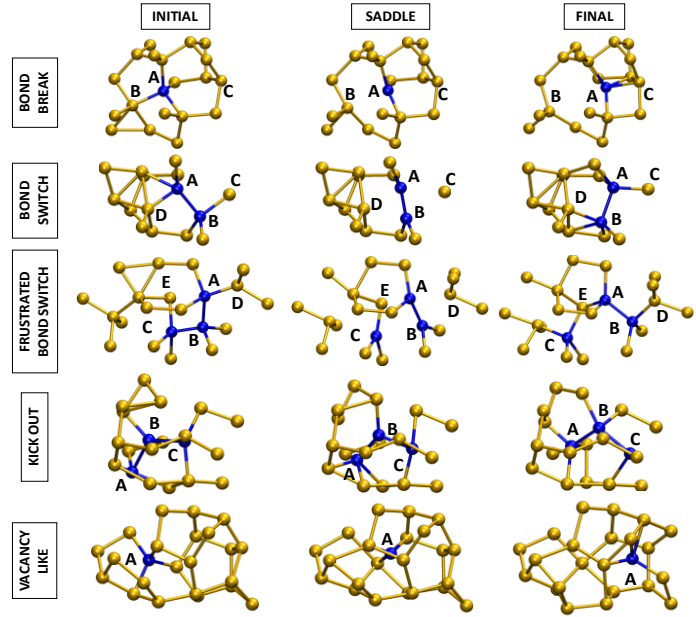

**Figure 2** (Color online) Snapshots of the observed atomic rearrangements. See text for details.

The BB only involves one atom (A) in the rearrangement. When one of its bonds is broken (bond A-B), the influence of the rest of its neighbors makes the atom A to move in the opposite direction to the broken bond. It forms a new bond with another atom (C) which was initially at a second neighbor distance. Atomistic models have proposed the BB as a possible mechanism for the creation of light induced defects in hydrogenated *a*-Si [16]. The BS consists of the interchange of two neighbors (C, D) between two neighboring atoms (A, B): the bonds A-D and B-C are switched to become A-C and B-D. While the BS mechanism was so far only guessed [15], in the present study by analyzing the TBMD trajectories we provide evidence that it indeed occurs and is responsible for large local rearrangements. Furthermore, our results also indicate that BS in *a*-Si can be produced by thermal atomic motions in addition to light illumination, as it was proposed by Wagner *et al.* [17].The BS has a special relevance in both *c*-Si and *a*-Si. The so-called `IV pair' is a point defect in *c*-Si that can be generated and recombined through a BS [6, 18]. This defect plays an important role in describing the amorphization and the recrystallization processes of Si [18]. In addition, the so-called *concerted exchange* diffusion mechanism in *c*-Si can be regarded as a combination of three consecutive BSs involving the same pair of atoms [19]. Moreover, the BS is the fundamental movement within the WWW bond network for the generation of *a*-Si from *c*-Si.

It is worth to note that, due to rearrangements induced in the surroundings, sometimes the neighbor interchange is not totally completed in the BS event. We have proved this feature through the identification of a frustrated BS (FBS) rearrangement, which shown in Figure 2. The bonds A-D and B-C break as it occurs in the BS, and the bond B-D is formed. However, atom A does not bond to atom C but rather to atom E, which is a different neighbor of atom B. Although at the end of the movement the new formed bonds are A-E and B-D, this type of rearrangement has a BS nature. During the KO event, one atom (A) displaces another one (B) from its position. As a consequence of this rearrangement, atom C is also displaced in this particular case. This kind of movement is similar to that occurring during the diffusion of Si self-interstitials in *c*-Si [6, 14, 20]. Finally, in the VL only one atom (A) moves as in the BB, but in the VL the distance traveled is longer. In the case of the VL, atom A seeks for open spaces in the network, and changes most of its neighbors during the displacement. This movement resembles the vacancy diffusion mechanism in *c*-Si where an atom moves towards the empty space associated to a vacancy [21].

Hence, similar atomic rearrangements exist in both *c*- and *a*-Si, which might be due to their similar short-range atomic environment. It is worth to note that although the BB and the BS events found here agree with the observations of previous theoretical studies [15], neither the KO nor the VL events were reported.

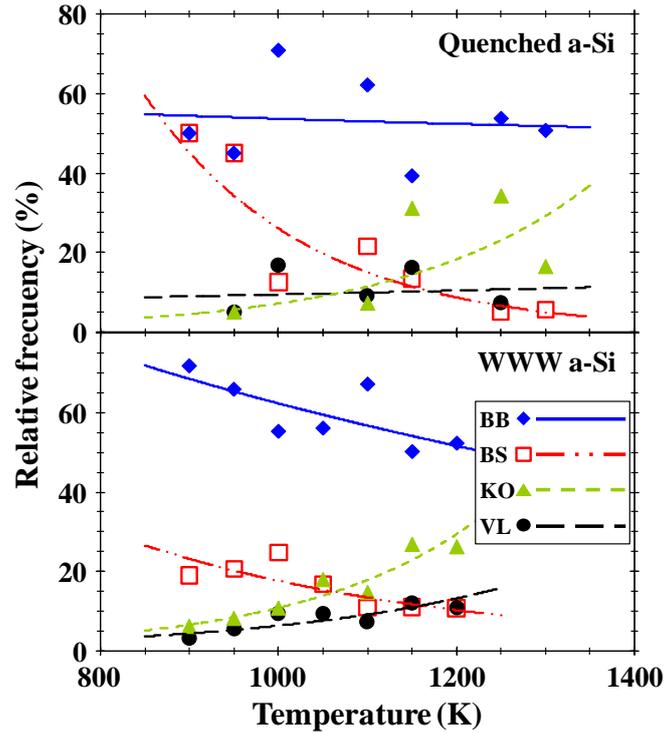

**Figure 3** (Color online) Statistics of the observed rearrangement mechanisms. Lines are to guide the eye.

In order to evaluate the relevance of the observed mechanisms, we have represented in Figure 3 the relative frequency of a given event, i.e. its relative percentage of occurrence, as a function of temperature. Since sometimes it is difficult to distinguish between the FBS from a BS, he have gatherer FBS events within the BS group. We have found that the BB rearrangement is the more frequent event in both samples at the analyzed temperatures. The BS has an important contribution to atomic rearrangements a low temperatures, especially in quenched *a*-Si, but its relevance diminishes as temperature increases. The relative frequency of the KO events increases with temperature, while in the case of VL the increase is softer. In general, it can be inferred from Figure 3 that coordination defects favors BS events, while KO and VL events take place when the temperature is high enough. All these rearrangements caused the steps in the MSD observed in Figure 1. Small steps at low temperatures are associated to isolated rearrangements involving small displacements of few atoms. As temperature increases, the higher steps are related to an increment of the complexity of rearrangements involving a larger number of atoms, which can be regarded as a succession of several `basic rearrangements' being usually the BB the triggering event.

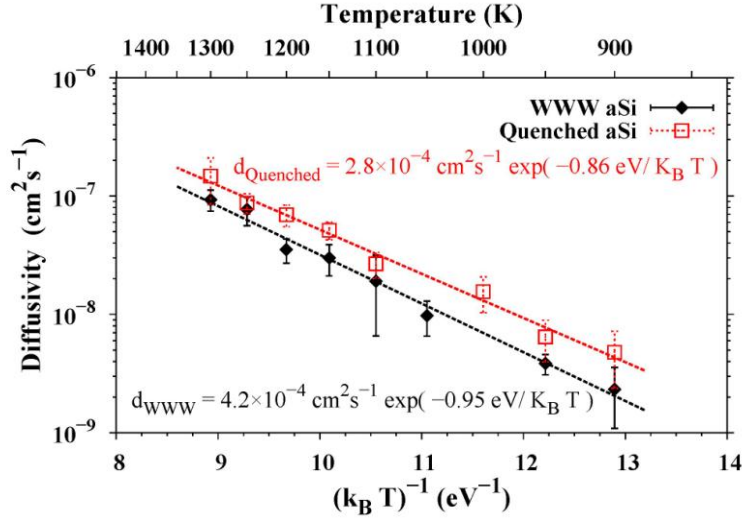

**Figure 4** (Color online) Average atomic diffusivities at each temperature of the samples considered. Straight lines and equations show the result of the Arrhenius data fit.

We have also calculated the atomic diffusivity $d(T)$ from the MSD at different temperatures using the Einstein relation

$$d(T) = \lim_{t \to \infty} \frac{MSD(t)}{6t}$$

**Equation 2**

In Figure 4 we represent the average diffusivity at each temperature for the samples considered. Quenched samples have higher atomic diffusitities than WWW ones. This may be due to their higher initial structural distortion, which appears to enhance atomic rearrangement events. Data can be described with an Arrhenius behavior within the temperature range analyzed:

$$d(T) = d_0 \exp\left(-\frac{E_m}{k_B T}\right)$$

**Equation 3**

where $d_0$ is the diffusivity prefactor, $E_m$ is the effective migration energy for self-diffusion in $a$-Si, $k_B$ is the Boltzmann constant, and $T$ is the temperature. Contrary to the case of $c$-Si where single self-diffusion mechanisms can be identified and studied separately to obtain their contribution to the overall atomic diffusion, this is not the case in $a$-Si since atomic rearrangements are not unique and well defined due to the underlying changing amorphous network. Instead, the effective migration energy calculated from Equation 3 captures the average behavior of the observed atomic rearrangement mechanisms. The result of the Arrhenius fits is shown in Figure 4. The obtained effective migration energies are 0.86 eV and 0.95 eV for quenched and WWW *a*-Si, respectively. It is worth to note that these energies are of the order of the migration energies of Si self-interstitials in *c*-Si [6, 14, 20]. As the effective migration energies obtained are very similar in both types of *a*-Si samples, it can be concluded that the underlying short-range atomic rearrangement mechanisms are mainly the same. Nevertheless, rearrangement events appear to occur more frequently in the quenched samples with respect to WWW ones according to the diffusivities represented in Figure 4.

In conclusion, we have studied from an atomistic point of view the microstructure evolution of *a*-Si during annealing using TBMD simulations. The evaluation of the MSD together with the analysis of atomic trajectories enabled us making a synopsis of the type and relevance of rearrangement events occurring during annealing. Despite the variety of mechanisms found, an effective migration energy of ~1 eV was obtained from the Arrhenius plot of the atomic diffusivities that encompasses their average behavior in self-diffusion in *a*-Si. This energy could be used in Monte Carlo simulations of the microstructure evolution of *a*-Si in order to access to time scales that go beyond molecular dynamics simulations.

Authors thank L. Bagolini for providing the WWW *a*-Si cells. This work was partially funded by HPC-EUROPA2 under project number 228398 with the support of the European Commission Capacities Area - Research Infrastructures Initiative, and by Spanish DGI under project number TEC2008-06069 and by Junta de Castilla y León under Project VA011A09 (IS, LAM, LP).

**REFERENCES**

* ivasan@tel.uva.es

[1] L. Khriachtchev, *Silicon Nanophotonics* (World Scientific Publishing, Singapore, 2008).

[2] S. Coffa *et. al.*, Phys. Rev. B **45**, 8255 (1992).

[3] S. Mirabella *et al.*, Phys. Rev. Lett. **100**, 155901 (2008).

[4] S. Roorda *et al.*, Phys. Rev. B **44**, 3702 (1991).

[5] I. Kwon *et al.*, Phys. Rev. B **49**, 7242 (1994).

[6] M. Tang *et al.*, Phys. Rev. B, **55**, 14279 (1997).

[7] L. Colombo, Annu. Rev. Mater. Res. **32**, 271 (2002).

[8] M. Cogoni *et al.*, Phys. Rev. B **71**, 121203 (2005).

[9] V. Rosato and M. Celino, J. Appl. Phys **86**, 6826 (1999).

[10] P. Biswas, Phys. Rev. B **65**, 125208 (2002).

[11] M. Ishimaru, S. Munetoh, and T. Motooka, Phys. Rev. B **56**, 15133 (1997).

[12] F. Wooten, K. Winer, and D. Weaire, Phys. Rev. Lett. **54**, 1392 (1985).

[13] W. C. Swope *et al.*, J. Chem. Phys. **76**, 637 (1982).

[14] L. A. Marqués *et al.*, Phys. Rev. B **71**, 085204 (2005).

[15] F. Valiquette and N. Mousseau, Phys. Rev. B **68**, 125209 (2003).

[16] M. Stutzmann, W. B. Jackson, and C. C. Tsai, Phys. Rev. B **32**, 23 (1985).

[17] L. K. Wagner and J. C. Grossman, Phys. Rev. Lett. **101**, 265501 (2008).

[18] L. A. Marqués *et al.*, Phys. Rev. Lett. **91**, 135504 (2003).

[19] K. C. Pandey, Phys. Rev. Lett. **57**, 2287 (1986).

[20] M. Posselt, F. Gao, and H. Bracht, Phys. Rev. B **78**, 035208 (2008).

[21] P. M. Fahey, P. B. Griffin, and J. D. Plummer, Rev. Mod. Phys. **61**, 290 (1989).